\documentclass[Journal]{IEEEtran}
\IEEEoverridecommandlockouts
\usepackage{cite}
\usepackage{amsmath,amssymb,amsfonts,nccmath,tabularx,booktabs}
\usepackage{graphicx}
\usepackage{float}
\usepackage{subfigure}
\usepackage{textcomp}
\usepackage{hyperref}
\usepackage{multicol}
\usepackage{graphicx}
\usepackage{xcolor}
\usepackage{svg}
\usepackage{algorithm}
\usepackage{algpseudocode}
\def\BibTeX{{\rm B\kern-.05em{\sc i\kern-.025em b}\kern-.08em
    T\kern-.1667em\lower.7ex\hbox{E}\kern-.125emX}}
\begin{document}

\title{fybrrLink: Efficient QoS-aware Routing in SDN enabled Next-Gen Satellite Networks\\
}

\author{
\IEEEauthorblockN{Prashant Kumar, Saksham Bhushan, Debajyoti Halder, and Anand M. Baswade}\\
\IEEEauthorblockA{Dept. of Electrical Engineering and Computer Science \\
Indian Institute of Technology Bhilai, India \\
Email: \{prashantk, sakshamb, debajyotih, anand\}@iitbhilai.ac.in}
}

\maketitle

\begin{abstract}
Providing high-speed Internet service using satellite network has attracted researchers from both academia and industry mainly due to the characteristics of Low Earth Orbit (LEO) satellite networks such as global coverage, scalability, and lower transmission delay. With the recent advancements in the Software-Defined Network (SDN), implementation of SDN in Non-Terrestrial Networks (NTN) can help to achieve the set goals for 5G and beyond networks. Since satellite networks have a distinct architecture, some of the traditional protocols no longer remain useful. Therefore, to satisfy the diverse Quality of Service (QoS) requirements for a variety of applications, we propose a novel and centralized QoS-aware routing algorithm, called \emph{fybrrLink} in which the global view of the network in SDN is utilized. We implement a modified Bresenham's algorithm and Dijkstra's algorithm to find the optimal path in a significantly reduced computation time. Also, taking advantage of the deterministic satellite constellation, we propose a flow rule transfer algorithm and a topology monitoring algorithm. Further, \emph{fybrrLink} is evaluated with multiple NS3 simulations, and results confirm its supremacy over other state-of-the-art algorithms.
\end{abstract}


\section{Introduction}
Due to the ever growing demand for connectivity and also the rapid increase in smart devices, there is a need for a more scalable and globally connected network. The concept of Non-Terrestrial Networks is deemed to be the perfect candidate for future networks, providing services to the uncovered geographical areas as it is costly to cover rough terrain with optical fibres. The existing optical fibre links are prone to natural calamities, thus their maintenance is a major concern. Since, satellites have wider coverage and scalability in comparison to terrestrial networks, it can be a major factor in covering the shortcomings of the terrestrial networks. Several big corporations such as SpaceX \cite{starlink} is developing a satellite based low latency, broadband internet system to meet the needs of consumers across the globe.

It is estimated that by the end of 2024, a smartphone will consume more than 21GB of data per month on average (nearly 4 times the amount consumed in 2018). In addition to this increased usage, the number of smartphone subscriptions is set to increase by 45 percent, reaching a total of 7.2 billion \cite{ericsson}. Thus meeting the needs of users with the required QoS with global reach becomes crucial and requires a scalable system such as NTN. Next-Gen space internet service is being built to improve the benchmark downlink and uplink speed of state-of-the-art internet services. Space snternet would help terrestrial networks to meet the needs of public demand. SpaceX, Amazon and OneWeb are among the few who have started plans to either launch satellites or have already set up satellites for testing their services. With around just 800 satellites deployed and active, Starlink claims downlink speed of upto 150 megabits per second and latency of 20 ms to 40~ms~\cite{starlinksite}. 

Software-Defined Network (SDN) is an approach that enables centralized and programmable control by separating control and data plane. Flexibility, programmability, and logical centralization are its main features with which network management is simplified and operational costs are reduced~\cite{6994333}. The incorporation of SDN in the non-terrestrial networks is in its early stage and there are a lot of issues to be handled before the actual implementation \cite{8258968}. We address some of the issues in this paper as frequent handovers and mobility of the satellites deems the existing terrestrial-based routing algorithms void.

Applications that require multimedia data transfer have their own QoS requirement which must be fulfilled to improve user experience and for efficient resource management. In comparison to the terrestrial networks, satellite architectures have much-limited computation and storage resources with never-ending handovers. Many of the recent researches focus only on finding the path with fewer hops but the shortest path might not be able to guarantee the required QoS \cite{8515722}. They allocate a fixed amount of bandwidth to each flow which is not adequate to fulfill the needs of QoS extensive applications such as online gaming, media streaming, etc, \cite{8567898}. There exist diverse QoS requirements as some services are bandwidth-sensitive while some are latency-sensitive \cite{qosIntro}. The routing algorithm should also consider congestion in the network as a factor while computing the path as it leads to more packet loss and more queuing delay. Global view and centralized control of SDN controller must be utilized to consider the current status of links during routing. Thus, the development and design of QoS-aware routing algorithms for satellite networks is very crucial for a better user experience.

The contributions of the paper can be summarized as follows:
\begin{itemize}
    \item We propose an efficient QoS aware routing algorithm for satellite networks which computes optimal path with a significant improvement in computation time when compared to other state-of-the-art routing algorithms for NTN.
    \item We propose a method for congestion avoidance during routing and quick re-routing during Inter-Satellite Link (ISL) congestion or satellite failure by incorporating a scoring function for each link.
    \item We also discuss first-of-a-kind flow-rule transfer algorithm for non-disruptive service during satellite handovers by using the predictable topology of the satellite constellation.
    \item We also provide an algorithm for topology discovery and monitoring to collect real-time network state information.
\end{itemize}

The rest of the paper is organised as follows. Related work is discussed in \hyperref[sec:relatedWork]{Section II}. In \hyperref[sec:systemModel]{Section III} and \hyperref[sec:protocols]{Section IV}, the proposed scheme and algorithms are discussed. In \hyperref[sec:performanceEvaluation]{Section~V}, fybrrLink is evaluated using NS3 simulator and the results and key findings are discussed. Finally, \hyperref[sec:conclusion]{Section VI} concludes the~work.

\section{Related Works}
\label{sec:relatedWork}
Traditional satellite networks are less flexible and they are highly dependent on closed and planned architecture which makes them unscalable. To cope with these issues, multiple SDN-based architectures are suggested. In \cite{8258968}, an SDN-enabled multi-layer controlled architecture with extended open-flow protocol is proposed to enhance the programmability and flexibility of the satellite network.
In \cite{8610424}, the authors talk about the space-ground integration. Another SDN-based architecture is proposed which will facilitate efficient network management, increase network flexibility, and provide an enhanced QoS to the users.
SAGECELL \cite{8436052} takes the element from space, air, and ground domains, including satellites, UAVs, vehicles, static macrocells, and small cells, gateways, routers, and so on to provide fast network reconfigurability, seamless interoperability, and adaptive network control.
In \cite{8258968,8610424,8436052}, Network hypervisors are utilizing the global view of the network to efficiently use the pooled resources by locating non-conflicting network resources among satellite network virtual operators. So satellite networks can be used as a service.

A heuristic ant colony-based end-to-end fragment-aware path calculation (ACO-EFPC) algorithm is presented in \cite{8567898} for an efficient bandwidth allocation, improved wavelength fragmentation, and better load balancing in Integrated Satellite Terrestrial Network (ISTN). Status Adaptive Based Dynamic Routing (SADR) \cite{7878015} improves Ant Colony Optimization (ACO) algorithm to find QoS optimal paths by increasing more pheromone concentration to the path with better QoS status value. 
 
Various link-state algorithms are also proposed for the routing purposes in LEO satellite networks\cite{9367496,8761611}. As their asynchronous link-state update approaches are distributed, a huge amount of link update messages will be flooded all over the network for global route convergence even during the frequent regular topology changes. Sway, an SDN based QoS-aware routing protocol for delay-sensitive and loss-sensitive application, is presented in \cite{8385144}. The greedy routing algorithm in Sway is based on Yen's K-shortest path algorithm \cite{yen}. The path satisfying the QoS requirements of the incoming flow is chosen among the top K shortest paths for forwarding purposes.
 
A software-defined routing algorithm (SDRA) for NTN is proposed in \cite{8253282} to obtain the optimal routes. Controllers can bypass the congested Inter-Satellite Links (ISLs) for the routing calculation. A master controller and several slave controllers are distributed over the earth to realize flexible centralized management of the satellite network. Software-defined space networking (SDSN)~\cite{15728153} includes a strategy-based routing algorithm that can be used for both SDN and non-SDN-based architecture. Even when inter-orbit links are not much stable, they are prioritizing them over intra-orbit links. None of the SDN-based routing algorithms for satellite networks \cite{8253282,15728153} differentiates between various traffics. They do not consider the network status for routing the internet traffic and thus, do not provide the required QoS.
 
\section{Proposed Architecture}
\label{sec:systemModel}
In this section, we present the base architecture and various control layer modules for our proposals. We start by explaining briefly the adapted SDN architecture upon which our proposed approach is based. We also propose the physical architecture and constraints of the deployed satellites. Then we explain various terms required for clear understanding of the proposed algorithms and their performance evaluation. We mention the three modules proposed in fybrrLink and explain the method of cost calculation for each communication link in the satellite network.
\subsection{Elements of SDN architecture}

NTN is a network spread across the globe and is expected to surpass the current day terrestrial day traffic. Therefore, a single controller cannot be deployed for such a vast system, as the computation and signalling delay at the controller will limit the scalability and hinder user experience. We follow a similar SDN-enabled NTN architecture consisting of data and control plane as proposed in SoftSpace \cite{8258968}. The architecture is divided into four segments, (i) Space segment consisting of satellites organized in the consellation. (ii) Ground segment consisting of satellite terminals (STs) for connecting user-end devices to the network and satellite gateways (SGWs) used for connecting satellites and network management entitites at ground. (iii) Control and management segment, this consists of real-time network management and control functionalities. (iv) User segment consisting of the user-end devices for various applications using the satellite network.
\begin{figure}[!htb]
    \centering
    \includegraphics[width=8cm]{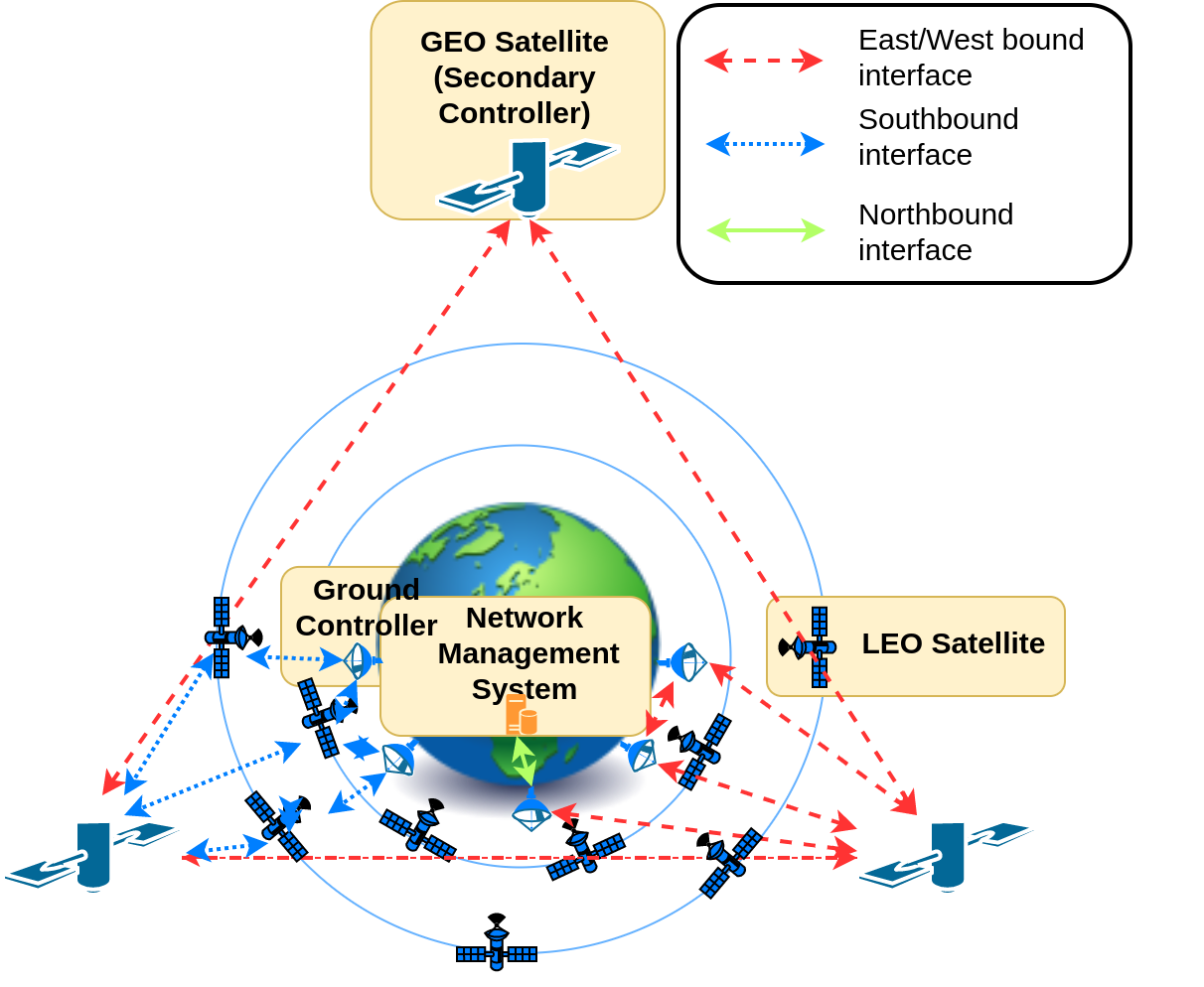}
    \caption{Elements of SDN architecture.}
    \label{fig:sdn_arch}
\end{figure}

Various elements of our physical architecture are depicted in Fig. \ref{fig:sdn_arch}. The data plane consists of the software-defined LEO satellites (SD-LEO), software-defined satellite gateways (SD-SGWs), software-defined satellite terminals (SD-STs) and the interconnecting software-defined switches. The control plane is the network brain consisting of network management and control entities of the network which is the ground controller and the Geostationary Orbit (GEO) controller satellites. Since, a single controller will have limited computation and will not be able to cope with the network expansion and limit the scalability. Also, the network will be vulnerable to single point failure at the controller. We follow a multi-layer controller architecture to overcome the problems with single controller architecture. We propose a two layered controller architecture consisting of the GEO controllers layer and the ground controllers layer. The GEO controller layer can also be termed as secondary controllers or domain controllers. The LEO satellites are grouped into various domains according to the coverage of GEO satellites and the GEO satellites are responsible for the flow control management of a particular domain. For example, a route within a domain will not require ground controllers to be involved for the route computation. Thus, each LEO satellite within a domain forwards its network state information to the domain controller. The domain controller then forwards the network state information of all the satellites in the domain to the ground controller for the ground controller to compile and compute the network-wide states. Ground controllers are the primary controllers responsible for inter-domain control instructions and routing. We are using several ground controllers to offer flexibility, programmability, and logical centralization in the network. Also, ground controllers generally have high storage and computational capability in comparison to GEO satellites.
A backhaul connection can also be introduced directly between the LEO satellites and ground controllers for fault detection and recovery mechanism, details of this are out of scope of this paper.
Further, extended OpenFlow protocol can be used for the communication between aforementioned planes. SDN controllers and network management system together can perform network control and resource management.

\subsection{Communication Links and Cost Calculation}
\label{subsec:score}
For the sake of simplicity, we are assuming that every orbital plane is parallel to each other. So at a point in time, there exist two intra-orbital links for each satellite and the others being diagonal links \emph{i.e.,} to the ones in different orbits. Two satellite nodes will have a link between them if they are visually connected \emph{i.e.,} the distance between them is not greater than a threshold distance. Fig. \ref{fig:dist_calc} shows an example scenario with two satellites X and Y. In order to be visible to each other, both satellites (satellite X and satellite Y) should satisfy the following equations:
 
\begin{figure}[t]
    \centering
    \includegraphics[width=8cm]{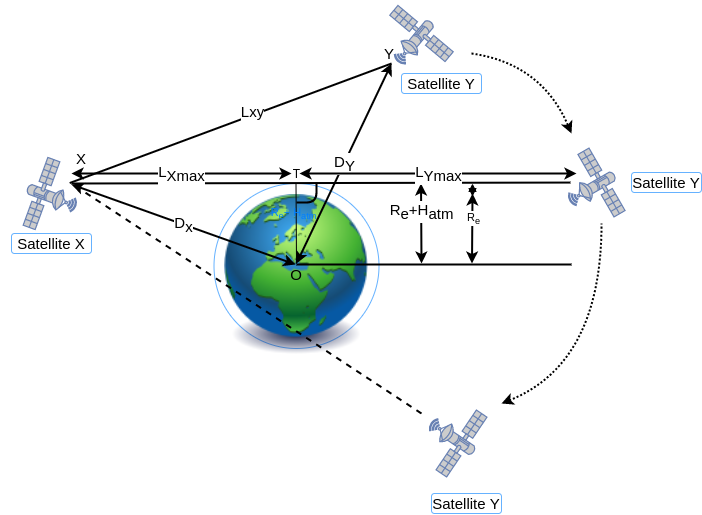}
    \caption{Diagram showing geometric visibility range of satellites with respect to each other.}
    \label{fig:dist_calc}
\end{figure}

\begin{align}
L_{XY} &< L_{Xmax} + L_{Ymax} \\
\text{where, }\nonumber\\
L_{Xmax} &= \sqrt{(D_{X})^{2} - (R_{e} + H_{atm})^{2}}\nonumber \\
L_{Ymax} &= \sqrt{(D_{Y})^{2} - (R_{e} + H_{atm})^{2}}\nonumber 
\end{align}

We have considered two types of Inter-Satellite Links in our network architecture:
\begin{enumerate}
    \item \textbf{Intra-Orbit Links:} Links between the adjacent satellites present in the same orbit.
    \item \textbf{Inter-Orbit Links:} Links between the satellites present in the adjacent orbital plane.
\end{enumerate}

\noindent
\textbf{Link Cost Calculation:} Let $G(V, E)[t]$ be the topology of the network at some instant t where V is the set of LEO satellites and E represents the set of links that exists between them. Every edge or link will be allocated a score and a cost after taking various factors into the account in order to generate QoS optimal routes. Those factors are listed below: 
\begin{itemize}
    \item $B_{ij}$ represents the bandwidth of $link_{ij}$ and specifies the maximum amount of data that may be sent over a certain link. 
    \item $Load_{ij}$ is a fraction of $B_{i,j}$ which is as of now getting utilized. Links that are part of multiple routes are busier and carry more traffic.
    \item $CD_{ij}$ is the congestion degree of the $link_{ij}$. We defined congestion as follows
        \begin{equation}
            CD_{ij} = \frac{Load_{ij}}{B_{ij}}
        \end{equation}
        A link with zero CD is considered idle, and forthcoming flows can be routed via it. Link with CD value as unity is essentially working at its full capacity and using it for any new flow will increase the queuing delay and packet loss rate of this link. Over congested links will have CD value greater than one and should be strictly avoided.
    \item $D_{ij}$  is the delay associated with the $link_{ij}$ and it is directly proportional to the distance between respective endpoints \emph{i.e.,} satellites.
        \begin{equation}
            D_{ij} = \frac{distance\_between\_two\_nodes}{speed\_of\_light}
        \end{equation}
    \item $PLR_{ij}$ is the packet loss ratio of the $link_{ij}$. Links with more packet loss rate are less reliable.
        \begin{equation}
            PLR_{ij} = \frac{PacketTransmitted_{ij}-PacketReceived_{ij}}{PacketTransmitted_{ij}}
        \end{equation}
    \item The portion of $B_{ij}$ that is still available to be utilised and can be used to route more flows is known as available bandwidth. $AB_{ij}$ represents the available bandwidth of the link, but is also dependent on $CD_{ij}$ in the following manner.
        \begin{itemize}
            \item If a link is congested \emph{i.e.,} $CD_{ij} > 0.8$:
                \begin{equation}
                    AB_{ij} = 0
                \end{equation}
            \item Else:
                \begin{equation}
                    AB_{ij} = B_{ij} - Load_{ij}
                \end{equation}
        \end{itemize}
    \item  Variation in the time delay between the signal transmission and reception is called Jitter. $J_{ij}$ is the average jitter associated with the $link_{ij}$ and can be computed using following equation:
        \begin{equation}
            \begin{split}
                & Jitter_{ij}(t) = \frac{1}{(t-startTime)}* \\
                & \sum_{k = startTime}^{t-1} (|Latency_{ij}(k) - Latency_{ij}(k+1)|)
            \end{split}
        \end{equation}
    \item $Stability\_Flag_{ij} = 1$ if it is a intra-orbit link and $0$ otherwise. The intra-orbit links always hold the connection since the relative position remains unchanged but inter-orbit link becomes intermittent along with the node movement.
\end{itemize}

The \textit{score} indicates how suitable a certain link is for a forthcoming flow. Above mentioned properties of the links are continuously monitored so that optimal routes can be found for every flow. The fitness score of each edge can be defined~as 

\begin{multline}
\label{eqn:score}
    Score_{ij}= k_{1}AB_{ij} + \frac{k_2}{latency_{ij}} + k_{3}(1- PLR_{ij})\\
    + \frac{k_4}{Jitter_{ij}} + k_{5}(Stability\_Flag_{ij})
\end{multline}
Apart from the properties of a link, Eqn. \eqref{eqn:score} also contains the multiple constants and their values can be decided based on the requirement of flow. A delay-sensitive flow cannot compromise with the latency so $k_{2}$ can be raised suitably. Loss-sensitive flows, on the other hand, necessitate a high $k_{3}$ value. So, the flexibility of adjusting the importance of parameters based on the requirement of flow is also provided.
Further, the \textit{cost} of the link will be defined as
\begin{equation}
    Cost_{ij} = \frac{1}{Score_{ij}}
\end{equation}
Links with a higher score will be preferred to include in final routes, since they are assigned smaller costs.

\textbf{Incoming flows:} A data flow will consist of the pair of source and destination vertices and the following application specific QoS requirements -
\begin{enumerate}
    \item Minimum required bandwidth $B$.
    \item Maximum allowed end to end delay $D$.
    \item Maximum allowed jitter $J$.
    \item Maximum allowed packet loss ratio $PLR$.
\end{enumerate}
Routing applications running inside the SDN controllers will generate QoS optimal paths for each incoming data flow. Further details about routing can be found in Section \ref{sec:routing}



\section{Proposed Approaches}
\label{sec:protocols}

In this section, we explain in detail the fybrrLink algorithms. We also describe the satellite constellation on which this proposed approach is based. We propose algorithms for the~3 modules on top of control layer - QoS aware routing and congestion handling, flow rule transferring, and topology discovery.

\begin{enumerate}
    \item \textbf{QoS aware routing}: To calculate the most optimal path for a flow, we propose a QoS-aware routing algorithm which uses a score based approach (as mentioned in Section \ref{subsec:score}) to route the incoming flow requests optimally.
    \item \textbf{Flow-rule transfer}: This module is responsible for assisting the network in handling topology alteration problems. The topology in satellite networks is predictable, so controllers can proactively determine the flow rules according to the position of the satellites in future topology.
    \item \textbf{Topology discovery and monitoring}: SDN enabled LEO satellites inform the controller about the network status so that the controllers can create a virtual topology. LEO satellites are required to perform two duties, one is to relay the data and the other to continuously update the controller about the network status. Any change in the parameters of the link such as bandwidth, delay, congestion degree, etc. (defined in Section \ref{metrics}) must be informed to the controller.
\end{enumerate}

The fybrrLink algorithm is based on an Iridium-like satellite constellation \cite{iridium} in which satellites are in a pole to pole orbit. fybrrLink proposes each satellite to have 6 ISLs which are of two types - 2 longitudinal ISLs and 4 diagonal ISLs, instead of 4 total ISL links in Iridium constellation. All satellite orbital planes are at same altitude from sea level. Satellite constellations play an important role in the suggestion of a QoS-aware routing algorithm.



\setlength{\extrarowheight}{1pt}
\begin{table}[H]
 \centering
 \caption{Algorithm Function definitions}
 \label{table:params}
\begin{tabular}{|p{2cm}|p{5cm}|}
\hline
Function             & Definition                                              \\ \hline \hline
XYCoordinate       & Returns cartesian coordinates of given (latitude, longitude) pair                                    \\ \hline
bresenham       & Runs Bresenham and returns shortest path \newline                                   \\ \hline
isSatisfyingPath       & Checks if given path satisfies QoS requirements                          \\ \hline
getSearchGraph      & Returns the ideal parallelogram for a path \newline                                                  \\ \hline
DijkstraAlgo   & Runs Dijkstra to find the best QoS satisfying path                                       \\ \hline
assignCost       & Computes scores of each ISL and assigns cost to graph edge                                                 \\ \hline
updateFlowTable             & Flow table at controller updated with new flow rules                                       \\ \hline
\end{tabular}
\end{table}

\subsection{QoS aware routing}
\label{sec:routing}

Few QoS aware routing approaches exist in Non-Terrestrial Networks that have been proposed for SDN-enabled architectures, but they fail to minimize the route finding time and hence achieve lower QoS satisfaction factor. Most algorithms use Dijkstra, run on the whole network, to find the shortest path. In fybrrLink, we decrease the search space for Dijkstra. fybrrLink finds the shortest path using modified Bresenham's routing algorithm and creates a shorter search space by selecting fewer satellites around the shortest path to run Dijkstra. In this way, the final route is the shortest path with increased QoS satisfaction factor. 

Algo. (\ref{Algo:Route}) is the fybrrLink algorithm to calculate the most optimal path for a flow. The algorithm in this module maximizes the number of QoS constraint satisfied flows. We use Bresenham’s line algorithm to find the perfect shortest path. All the satellites in orbit would comprise the cartesian plane on which the algorithm would run. The plane can be imagined as the 2D representation of the Iridium constellation. 

The algorithm gives us a path consisting of $M$ diagonal and $N$ longitudinal ISLs which would be the shortest path from source to destination satellite. We can form a parallelogram of order $M \times N$ with the source and destination satellites at the two opposite vertices of the parallelogram, such that all paths between source and destination satellites with $M$ diagonal ISLs and $N$ longitudinal ISLs inside the parallelogram, will have equal length and will be the shortest paths. We call this the \emph{Ideal Parallelogram}. We run a modified Dijkstra algorithm inside the Ideal Parallelogram with the computed cost as ISL weights to find the current optimal path between source and destination satellites which would give the best Quality of Service measures.

\begin{algorithm}
\caption{\emph{fybrrLink} Routing algorithm}
\label{Algo:Route}
\begin{algorithmic}[1]

\Procedure{Routing}{$G,F, \text{max\_CD}, \text{delta\_CD} $}
    \For{all flow(f) in the Flow$\_$Set(F)}
        \State source ← G.XYCoordinate(f.source)
		\State dest ← G.XYCoordinate(f.dest)
		\State m,n,\text{initial\_shortest\_path} ← bresenham(G, source, dest) 
		\If {f.QoSenabled == False or isSatisfyingPath(\text{initial\_shortest\_path}, \text{f.flow\_constraints}) }
            \State updateFlowTable(f, \text{initial\_shortest\_path})
        \Else
			\State \text{search\_graph} = getSearchGraph(G, source, dest, m, n)
			\State \text{no\_of\_iter} ← $\frac{1-\text{max\_CD}}{\text{delta\_CD}}$
			\State isSuitablePathFound ← False
			\For{i in range(\text{no\_of\_iter})}
				\State path ← DijkstraAlgo(\text{search\_graph}, source, dest)
				\If{ isSatisfyingPath(path, \text{f.constraints})}
					\State updateFlowTable(f, path)
					\State isSuitablePathFound ← True
					\break
				\Else
					\State \text{max\_CD} ← \text{max\_CD} +  \text{delta\_CD}
					\State assignCost(Graph, CD, source)
                \EndIf
            \EndFor
			\If{isSuitablePathFound $==$ False }
				\State informSender(``Message":``Given Flow cannot be satisfied. Please relax some contraints", ``BestPath" : path)
			\EndIf
        \EndIf
    \EndFor
		
\EndProcedure

\end{algorithmic}
\end{algorithm}

\textbf{Edge Case - \emph{N=0 OR M=0:}} When the shortest path between source and destination satellites comprises of longitudinal ISLs or diagonal links only. We cannot form an Ideal Parallelogram in this case as it is the only shortest path and only the path output from Bresenham's algorithm will be treated as the final path. In case of congestion in any of these paths, we would deviate the route in the different possible directions and run Bresenham from a new source satellite to the destination to check which new route gives the shortest path to the destination and whether it is free of congestion as well.

\begin{figure}
    \minipage{0.23\textwidth}

    \includegraphics[width=4.3cm]{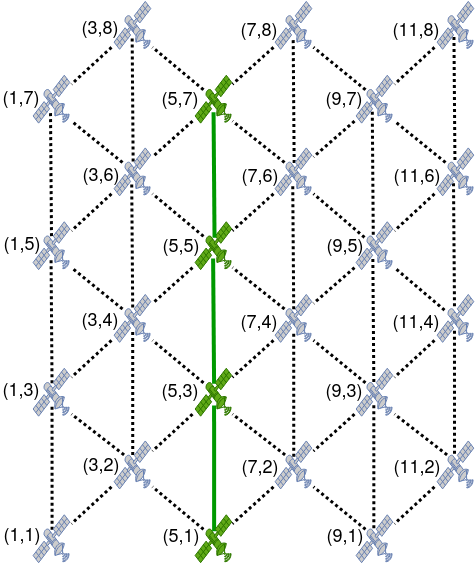}
    \caption{Before congestion.}
    \label{fig:bef_cong}
    \endminipage\hfill
	~
	    \minipage{0.23\textwidth}
       \includegraphics[width=4.3cm]{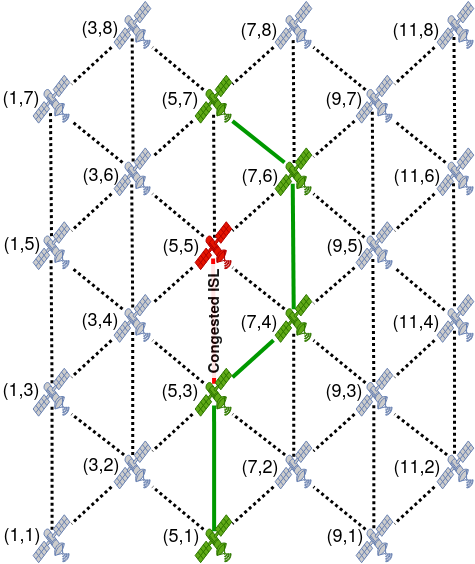}
    \caption{After congestion.}
    \label{fig:after_cong}
    \endminipage\hfill
\end{figure}

As shown in Fig. \ref{fig:bef_cong} the ISL [(5,3)-(5,5)] gets congested. We then calculate another route by running modified Bresenham between (5,3) and destination (5,7). The algorithm chooses the shortest and the least congested route, here, through (7,4) as shown in Fig. \ref{fig:after_cong}. Thereby, we are able to re-route the flow with low delay as well as evenly distribute the traffic in the~network.

\textbf{ISL Congestion and failures:} Normally if an ISL has traffic more than threshold ($CD_{ij} > 0.8$) then the cost of the ISL will be increased to a constant MAX value. This will avoid routing of a packet through that ISL and prevent ISL from being fully occupied which may also result in link failure. However if the ISL is congested due to satellite failure then we deliberately increase the cost of the ISL to MAX and re-route the existing routes to avoid that ISL. The congested/failed ISL is not in service unless the procedure for re-establishment of link is followed (beyond the scope of this paper).

\textbf{Note:} ISL congestions caused by satellite failures will deem all the 6 ISLs connected with the failed satellite to fail.

\subsection{Pre-defined flow rules}
In terrestrial networks with SDN there are no circumstances in which flow rules are to be transferred. However in non-terrestrial networks, states of a satellite constellation can be computed mathematically well in advance. This knowledge of future constellation state can be used to update flow rules in the satellites. Flow rule transfer module of fybrrLink is based on the same.

This module is responsible for assisting the satellite architecture in handling topology alteration problems. As per the predictable nature of the topology in satellite networks, controllers can proactively determine the flow rules according to the position of the satellites in future topology. We need not re-compute the routes after the handovers. This will avoid the interruption of ongoing services and expensive resources can be used for other purposes. Sometimes, the same flow will be routed with various paths at different instances of time. Hence, flow tables on the satellites must be updated according to the varying network topology.

The flow rule updation algorithm (Algo. (\ref{Algo:flow})) runs on a separate thread to maintain a seamless routing service. Satellites follow a predefined periodic trajectory around the Earth. The position of a particular satellite can be easily precomputed. Since, all the satellites in an orbital plane follow the same trajectory, they tend to take the place of the preceding satellite after a given time. The SDN controller notifies the current satellite at a position to update the flow rules in the new satellite prior to the handover. This ensures a smooth satellite handover and an uninterrupted service. 

\begin{algorithm}[!htb]
\caption{Flow Rule Transfer algorithm}
\label{Algo:flow}
\begin{algorithmic}[1]

\Procedure{setFlowRules}{$S, f$}
    \If{not firstSatelliteInPlane($S$)}
        \State f' ← getFlowRules($S$)
        \State setFlowRules(S.next, f')
    \EndIf
    \State S.flow = f
\EndProcedure

\Procedure{transferFlowRules}{$G$}
    \For{all orbitals(Orb) in G}
        \State S ← getFirstSatellite(Orb)
        \State f ← getFlowRules($S$)

        \State setFlowRules(S.next, f)
    \EndFor
\EndProcedure

\end{algorithmic}
\end{algorithm}

\subsection{Topology discovery and monitoring}
Topology discovery is a critical component of any SDN architecture. In order for the SDN controller to efficiently configure and manage the network, it needs to have up-to-date information about the state of the network and for this continuous network monitoring is required. There is no need to implement extra set of devices for this purpose. Here again fybrrLink uses the advantage of predictability of satellite constellation states to propose a more efficient topology discovery algorithm by just checking for satellite failures. In most of the existing terrestrial networks base stations are immobile, even if they are mobile, the mobility prediction cannot be done with precision, thus such an algorithm cannot be implemented for traditional networks.


Since, the satellites move on a pre-determined path with a constant velocity, the relative positions of the satellites can be calculated well in advance. This can be used for determining the topology of the satellite network at any given instance. Also, the satellite network will have a periodicity after which the topology will repeat. For that we can assume a satellite to have $p$ sets of neighbouring satellites in a period. Since, all the satellites are assumed to be moving identically, all the satellites have $p$ sets of neighbouring satellites. We can represent the $p$ sets with a state machine with $p$ states and hence each of the satellite has $p$ states, we can say that the whole network topology has $p$ states and after the $p^{th}$ state the topology returns back to the $1^{st}$ state. Therefore, the network topology has in total $p$ different configurations which can be pre-calculated at the controller. Within each state the LEO satellites are required to perform two duties, one is to relay the data and the other to continuously update the controller about the network status. Real-time updating is crucial because we are considering the bottleneck bandwidth and congestion of each link in our routing algorithm (See Algo. (\ref{Algo:Route}) for more details). If any link is broken, controller must be aware of that so that it is able to broadcast this information to other satellites and secondary controllers.

\begin{algorithm}[!htb]
\caption{Topology Discovery and Monitoring algorithm}
\label{Algo:topo-discovery}
\begin{algorithmic}[1]

\Procedure{sendPing}{S}       
   \State S.send({\lq heartbeat\rq: \lq alive\rq, \lq list\rq: S.adjList, \lq id\rq: S.id})
\EndProcedure

\Procedure{getPing}{msg}       
    \State adjList ← msg.list
    \State Sat ← msg.id
    \If{generateList(Sat) != adjList} \Comment{ISL failure}
        \State updateNetwork()
    \Else
        \Comment{continue with pre-computed network topology}
    \EndIf
\EndProcedure
\end{algorithmic}
\end{algorithm}

Referring to Algo. (\ref{Algo:topo-discovery}), every satellite will send a ping message to the ground satellite gateway along with the current adjacency list. Controller pre-computes the network topology mathematically, since the relative positions of each satellite can be estimated well in advance. If adjacency list of satellite does not match with the pre-computed list then there is a possibility of ISL failure due to congestion, else the pre-computed topology is used further for routing.
Referring to Algo. (\ref{Algo:topo-discovery}), every satellite will send a ping message to the ground satellite gateway along with the current adjacency list. Controller pre-computes the network topology mathematically. If adjacency list of satellite does not match with pre-computed list then it indicates that there is a congestion or one or more than one ISLs have failed, else the pre-computed topology is used further for routing.

\section{Performance Evaluation}
\label{sec:performanceEvaluation}
 
We simulated our routing algorithm and other benchmark algorithms with the help of NS 3.30.1 simulator and some packages of python such as networkx, matplotlib, etc. were used for complexity analysis and plotting purposes. Performance metrices on which we have evaluated fybrrLink are mentioned in the following subsection.

\subsection{Performance Metrics}\label{metrics}
We define performance evaluation metrics as
\begin{itemize}
    \item[-] \textbf{Average Routing Time} - It is the time taken by respective algorithms to compute the final route between the source and destination satellites for the data packet to flow.
    \item[-] \textbf{Packet Loss Rate (PLR)} - Packet loss rate is the ratio of number of packets lost and the total packets transmitted. It is directly proportional to the congestion in the network.
    \item[-] \textbf{Average Latency/End-to-end Delay} - The time taken for a packet to transmit from source to destination is end-to-end delay, which is a function of the path length and the congestion in the system.
    \item[-]\textbf{Congestion Degree} - It is defined as the ratio of used bandwidth and total bandwidth for a particular ISL. Higher congestion degree for an ISL means that the ISL is congested.
    \item[-]\textbf{Satisfaction Ratio} - It is the ratio of the QoS satisfied flows to the total number of incoming flows. Higher satisfaction ratio for an algorithm implies that the path generated by it is more optimal.
\end{itemize}
\subsection{Experimental Setup}
\label{Sub:setup}
We have used the NS3 network simulator for evaluating the performance metrics for this scheme as well as other schemes for comparison. We have also simulated the algorithms of the different schemes in Python to analyse and compare their time complexities.

\setlength{\extrarowheight}{1pt}
\begin{table}[!htb]
 \centering
 \caption{Simulation Parameters}
 \label{table:params}
\begin{tabular}{|p{4.5cm}|p{2.5cm}|}
\hline
Parameter             & Values                                              \\ \hline \hline
Number of test cases (flows)       & 324                                    \\ \hline
Simulation Time       & 324  seconds                                        \\ \hline
Incoming flow frequency       & 1 flow per second                           \\ \hline
Simulator      & NS 3.30.1                                                  \\ \hline
Number of satellites per orbit   & 10                                       \\ \hline
Number of orbits       & 10                                                 \\ \hline
Orbit Altitude             & 2000 Km                                        \\ \hline
maxCD (for Algo. (\ref{Algo:Route}))      & 0.82                                                          \\ \hline
deltaCD (for Algo. (\ref{Algo:Route}))       & 0.05                                                        \\ \hline
Allotted bandwidth for Intra-Orbit link      & 4.16 Gbps                    \\ \hline
Allotted bandwidth for Inter-Orbit link      & (3.10 - 3.70) Gbps           \\ \hline
Initial Congestion degree      &       0.60 - 0.80                          \\ \hline
Initial Packet loss rate & 0.0001 - 0.002                                   \\ \hline
Latency of Intra-Orbit links (at t = 0) & 17.5592 ms                        \\ \hline
Latency of Inter-Orbit links (at t = 0) & (2.7029 - 19.3630) ms             \\ \hline
[$k_{1}, k_{2}, k_{3}, k_{4}, k_{5}$]          &  [0.55, 0.30, 0.15, 0, 0]  \\ \hline
\end{tabular}
\end{table}
Table \ref{table:params} defines the parameters and values considered for the simulation. A total of 324 data flows were routed over a satellite network of 100 LEO Satellites. The LEO satellite mobility model of NS3 was allowed to assign the positions and velocities of each satellite at every instant. A new flow was requesting an optimal route from SDN controllers every second, and once routed, this data flow remained in the system until the simulation ended. The latency of every ISL is calculated from their length which can be easily obtained from the mobility model. All intra-orbit links were having the same distance (thus same latency too) of 5264.115 km because satellites were uniformly distributed in each orbit by the NS3 model. The latency of inter-orbit links was also allotted as per the positions of their respective end satellites only. Values of constants of edge scoring function were assigned in a way so that multiple ISL parameters such as bandwidth, latency, packet loss rate can be considered during routing.
We will be comparing fybrrLink with the multiple SDN based routing schemes in order to confirm advantages of our algorithm over others. 

\begin{figure}[!htb]
    \centering
    \includegraphics[width=7cm]{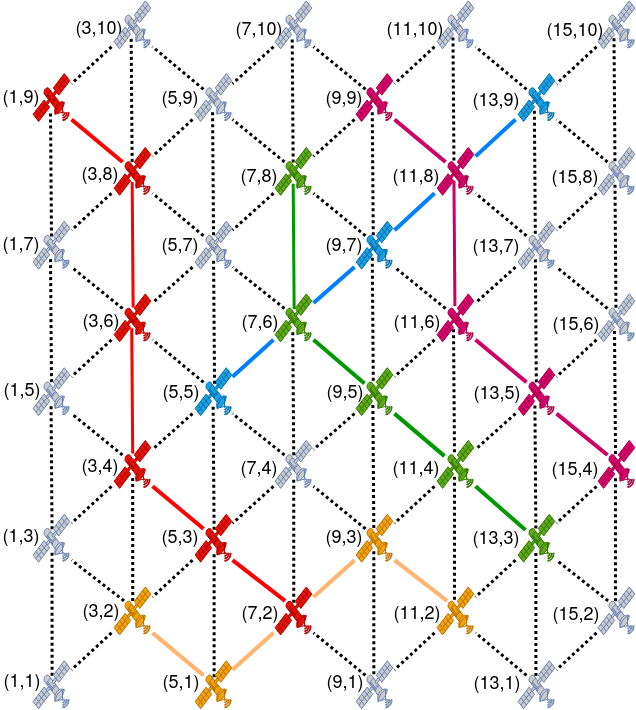}
    \caption{A part of the iridium constellation with flows indicating satellite routes computed with the modified Bresenham's routing algorithm.}
    \label{fig:iridium1}
\end{figure}

Fig. \ref{fig:iridium1} shows the graph representation of a part of the Iridium constellation with the vertices as the satellites and the edges as the ISLs between the satellites. The routes are being calculated between the source and destination satellites using the modified Bresenham routing algorithm. The colored nodes represent the intermediate satellites involved in the route for the transfer of the data packets.
\begin{figure}[!htb]
    \centering
    \includegraphics[width=7cm]{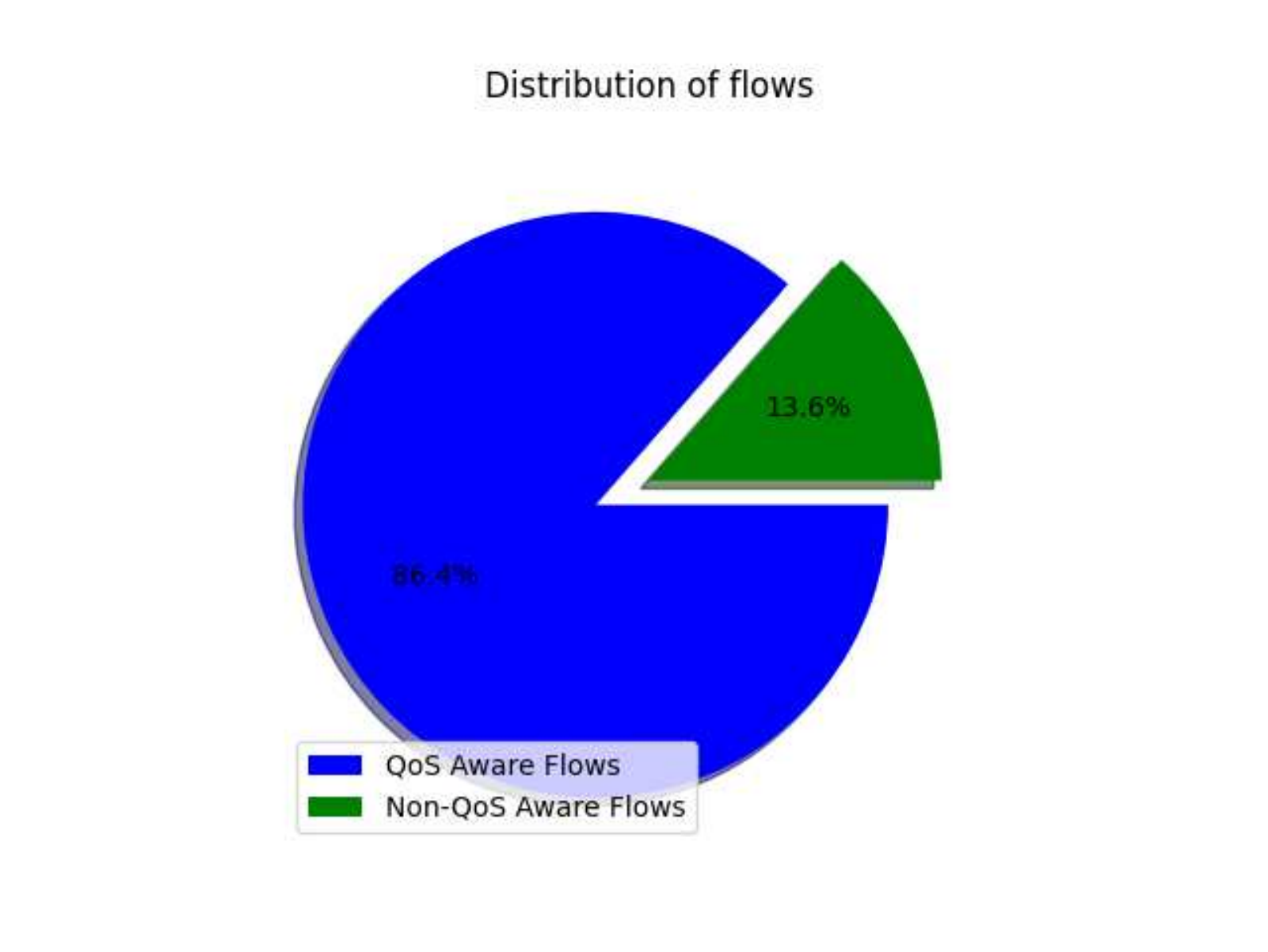}
    \caption{Distribution of QoS and Non-QoS aware flows during simulation.}
    \label{fig:distr_flows}
\end{figure}

We used a total of 324 test cases (flows) to evaluate our algorithm. As per the distribution in Fig. \ref{fig:distr_flows}, 280 flows was having certain QoS requirements but 44 flows was not having QoS requirement. Background traffic is one example for non-QoS aware flows.
We have compared proposed fybrrLink with following schemes.
\begin{itemize}
    \item \textbf{Dijkstra algorithm \cite{dijk}:} It is one of the most popular path finding algorithms used. It has been used for comparison in the maximum number of papers that propose routing algorithms. It is noteworthy that Dijkstra was not allowed to choose a congested path even if it possesses the lowest latency.
    \item \textbf{SDRA \cite{8253282}:} Software Defined Routing Algorithm in LEO Satellite Networks. This paper also proposes a routing algorithm, and various parameters which have been improvised in our paper.
    \item \textbf{Sway \cite{8385144}:} It is a traffic aware QOS routing scheme based on Yen's K-shortest paths algorithm for software defined IoT. Implementation of Yen's K-shortest paths algorithm is present in \cite{kshortestrepo}.

\end{itemize}

\subsection{Complexity Analysis}
Let $V$ be the total number of satellites in the constellation and $E$ be the total number of ISLs, then the time complexity for respective routing algorithms can be given as follows,

\noindent
Time Complexity of Dijkstra Algorithm = $O(V + E\log V)$.

\noindent
Time Complexity of K-Shortest Path Algorithm = $O(KV(E+V\log V))$.

\noindent
Time Complexity of All Paths DFS = $ O((2^V)(V+E))$.

\noindent
For fybrrLink algorithm, modified Bresenham's routing algorithm takes $O(N)$ time for computing the route, where N is the number of ISLs in the final computed route and $N$ is defined as $N = m + n$ where $m$ and $n$ are number of longitudinal and diagonal ISLs respectively required to reach destination node from source node.

\noindent
After performing Dijkstra inside the ideal parallelogram, the total time complexity will be $O(N) + O(V' + E' \log V')$ where 
\begin{align*}
    V' = \;&(m + 1)(n + 1)  \\
    E' = \;&(m + 1)n + (n + 1)m 
\end{align*}
\begin{figure}[H]
    \centering
    \includegraphics[width=9cm]{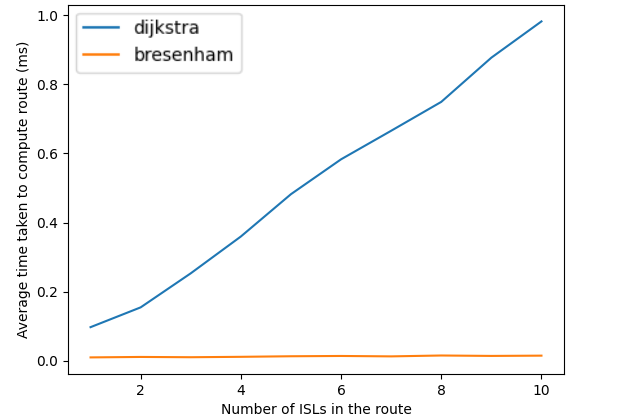}
    \caption{Execution time comparison of Shortest Path algorithms dijkstra and fybrrLink.}
    \label{fig:nonQoS_time}
\end{figure}

Fig. \ref{fig:nonQoS_time} shows the comparison of the average execution time of shortest path finding algorithms used in the schemes fybrrLink and Dijkstra for 350 flows. In fybrrLink, we use a modified Bresenham algorithm to find the shortest path in nominal time compared to Dijkstra.

We have also developed a visualization tool (fybrrLink-VizTool) that demonstrates the difference between the path computed using Dijkstra and modified Bresenham algorithm. It also shows the time taken by each algorithm to compute the path. The tool and it's source code is publicly available on \url{https://fybrrlink-viztool.herokuapp.com}.



\subsection{Experimental Results}
In this subsection, we showcase the experimental results which were obtained via NS3 simulation for the mentioned performance metrices. We compare our approach with 3 other centralized pathfinding algorithms. Some simulations and visualizations were also done using networkx library in python.

\begin{figure}
    \centering
    \includegraphics[width=9cm]{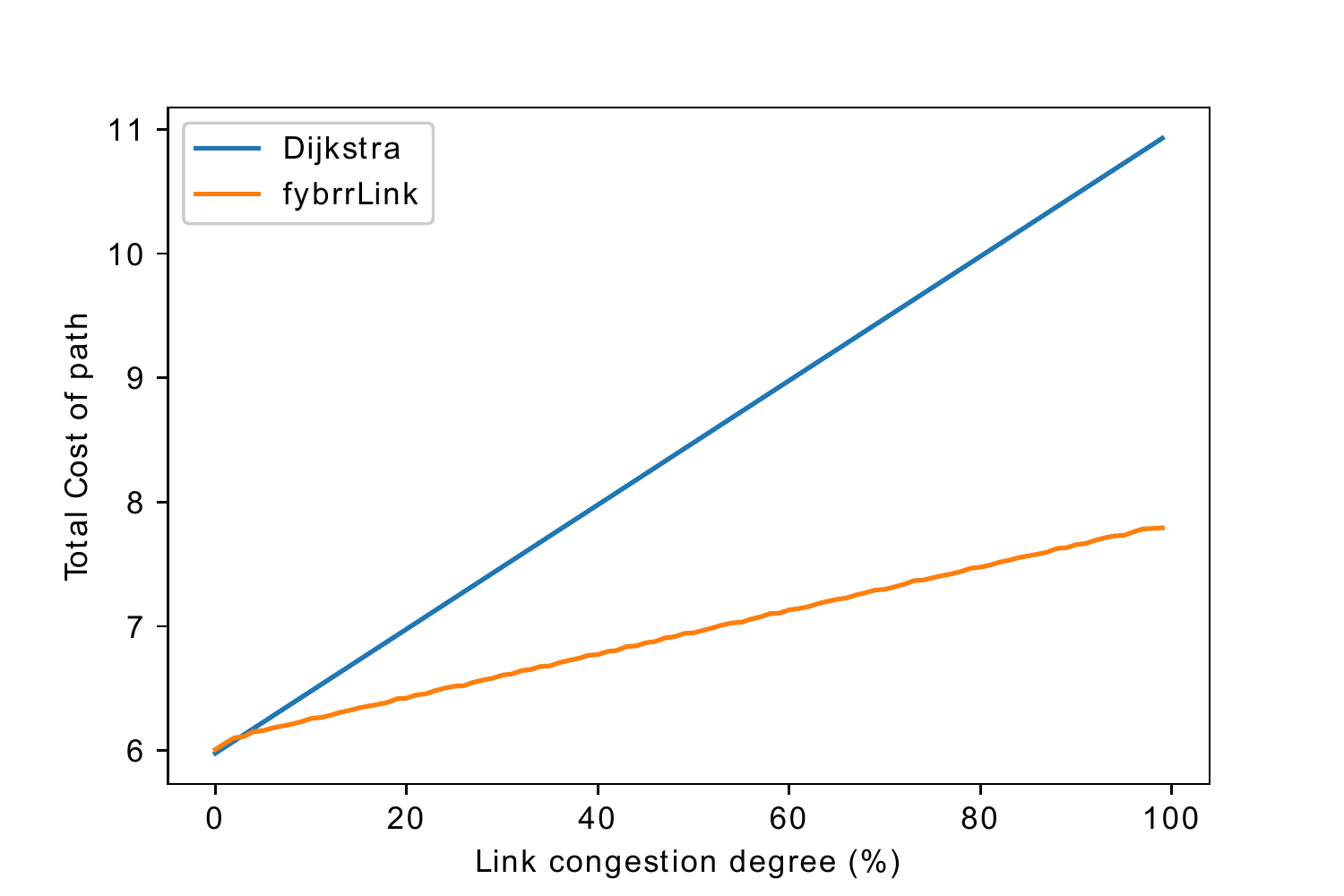}
    \caption{Total cost of the path with increasing load through the route.}
    \label{fig:loadvspkt}
\end{figure}
In Fig. \ref{fig:loadvspkt}, the total cost of the path between a pair of source and destination is plotted with increasing congestion on the ISL path for our approach as well as the Dijkstra algorithm. As the load increases in ISLs, our approach computes the cost for the ISL for every transmission and if the cost of the composite path increases beyond that of any other optimal path between the source and destination, then our approach finds the new optimal path and reroutes the traffic through the new path. Whereas the Dijkstra algorithm does not compute cost after transmissions, therefore after crossing certain threshold traffic, the path between the chosen source and destination gets fully congested. Therefore, it is evident that our approach distributes the traffic throughout the network in a more efficient and uniform manner, thus, decreasing the probability of congestion in the network. As can be seen from the plot that the cost of the path in our approach is $27\%$ lower than that in Dijkstra.

\begin{figure}
    \centering
    \includegraphics[width=9cm]{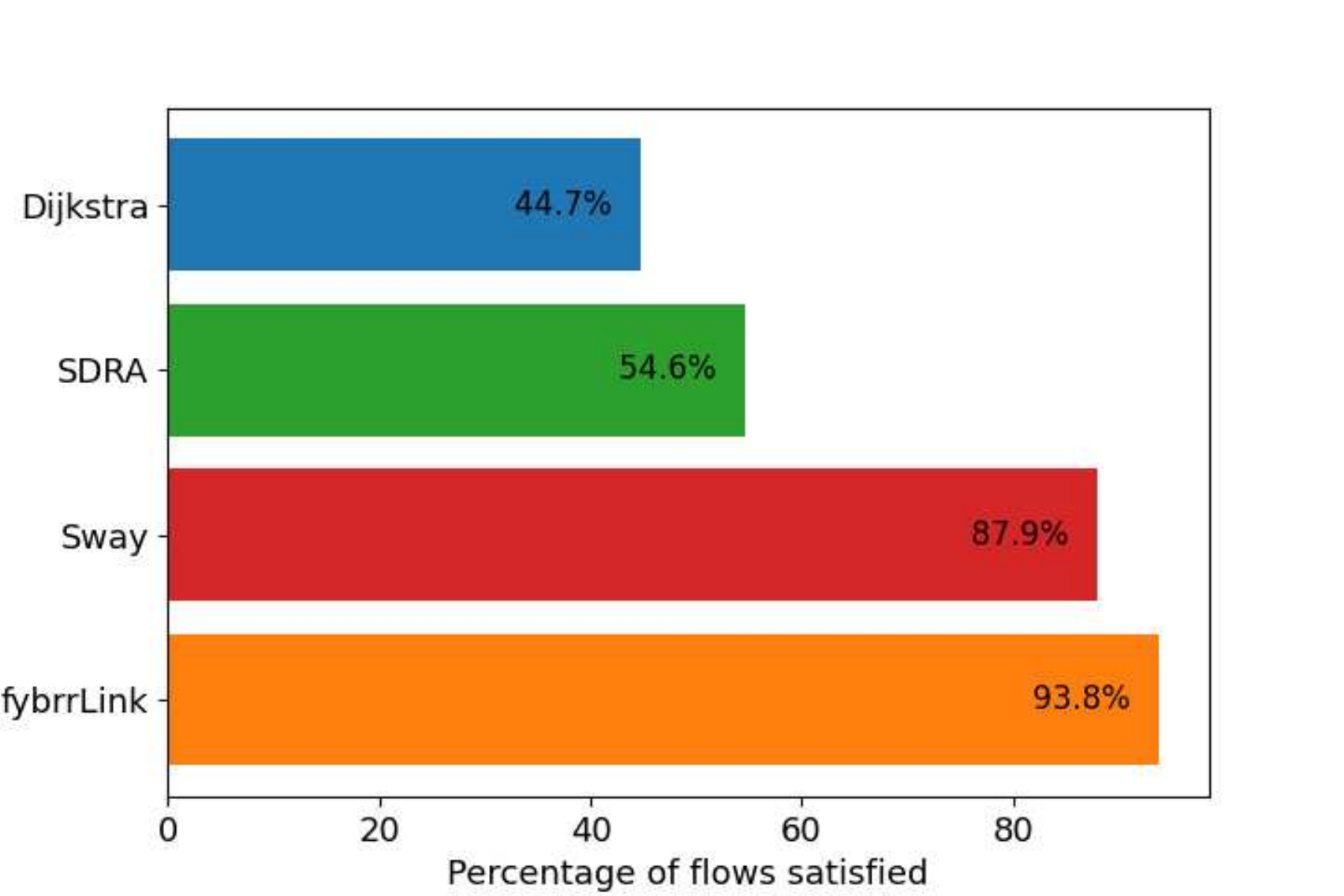}
    \caption{Percentage of flows satisfied by different algorithms.}
    \label{fig:dist_satisfaction}
\end{figure}

Fig. \ref{fig:dist_satisfaction} shows the supremacy of fybrrLink in finding the optimal QoS paths over other benchmark algorithms. Out of 324 flows, the QoS requirements of 304 flows were satisfied by fybrrLink but Dijkstra was able to fulfill the QoS needs of 145 flows only. Sway and SDRA satisfied 285 and 177 flows, respectively. Our algorithm satisfies more than double the number of flows satisfied by Dijkstra and produces more QoS aware paths in comparison to other schemes.

\begin{figure}
    \centering
    \includegraphics[width=8cm]{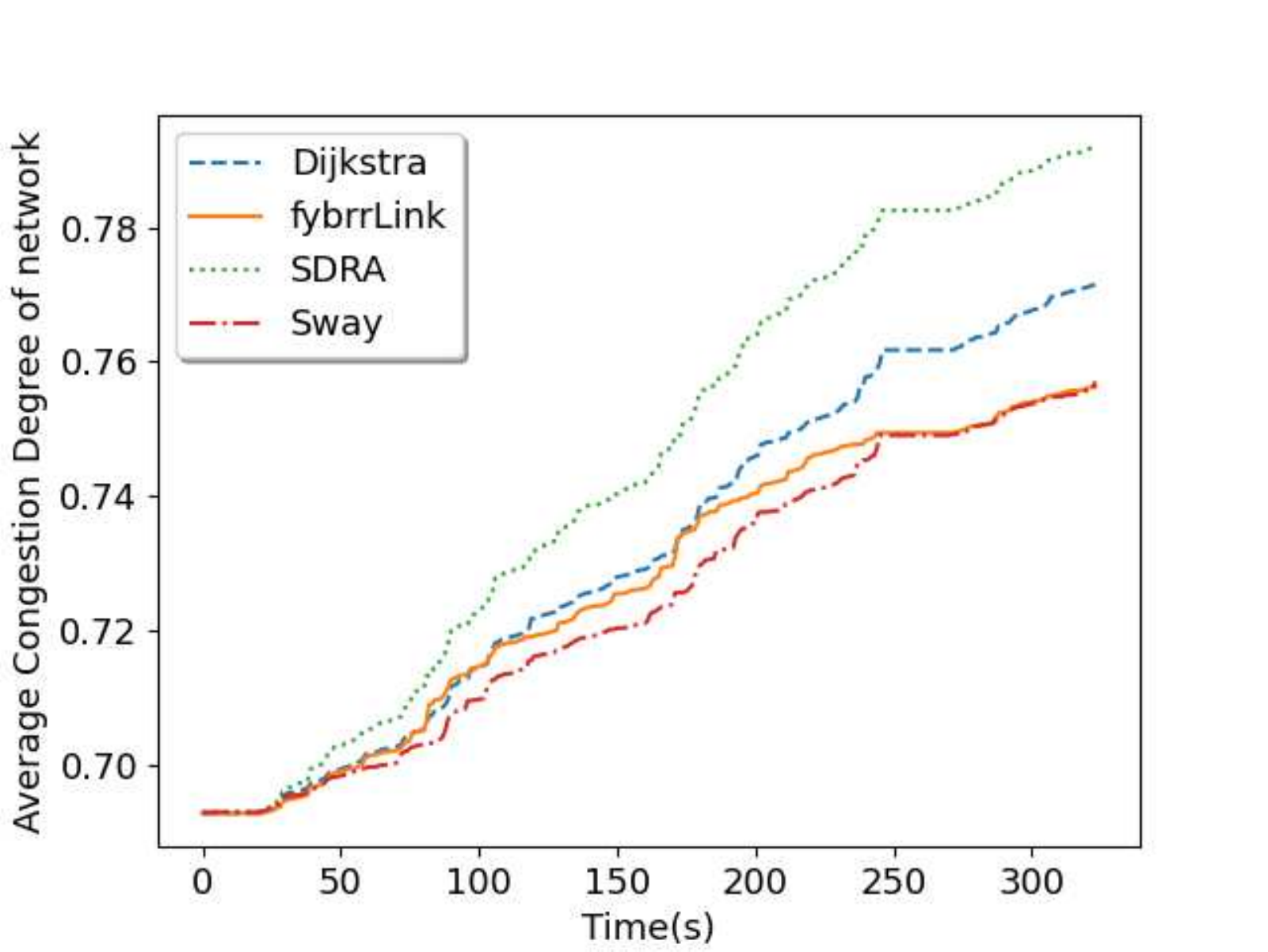}
    \caption{Average congestion of the network.}
    \label{fig:avgCD}
\end{figure}

In Fig. \ref{fig:avgCD}, congestion degree of all the ISLs are averaged and then the variation of this average congestion degree is being plotted with time. We can clearly observe that \emph{fybrrLink} algorithm is much better in load distribution among the ISLs. It makes sure that no link is getting overloaded by excluding the links that are going to be congested in the future from the routing calculations. We are also simulating Dijkstra with a congestion avoidance mechanism in which the congested paths are assigned the highest weights. SDRA even with its own congestion control mechanism is having a high average congestion degree because of its 4 link model. In Sway, the least congested and better QoS providing path is selected from the k shortest paths. Average (over time) of the average congestion degree of Sway and fybrrLink is $72.43\%$ and $72.72\%$, respectively, which is very much comparable to each other but values for Dijkstra and SDRA are $73.02\%$ and $74.49\%$, respectively.

\begin{figure}
    \centering
    \includegraphics[width=8cm]{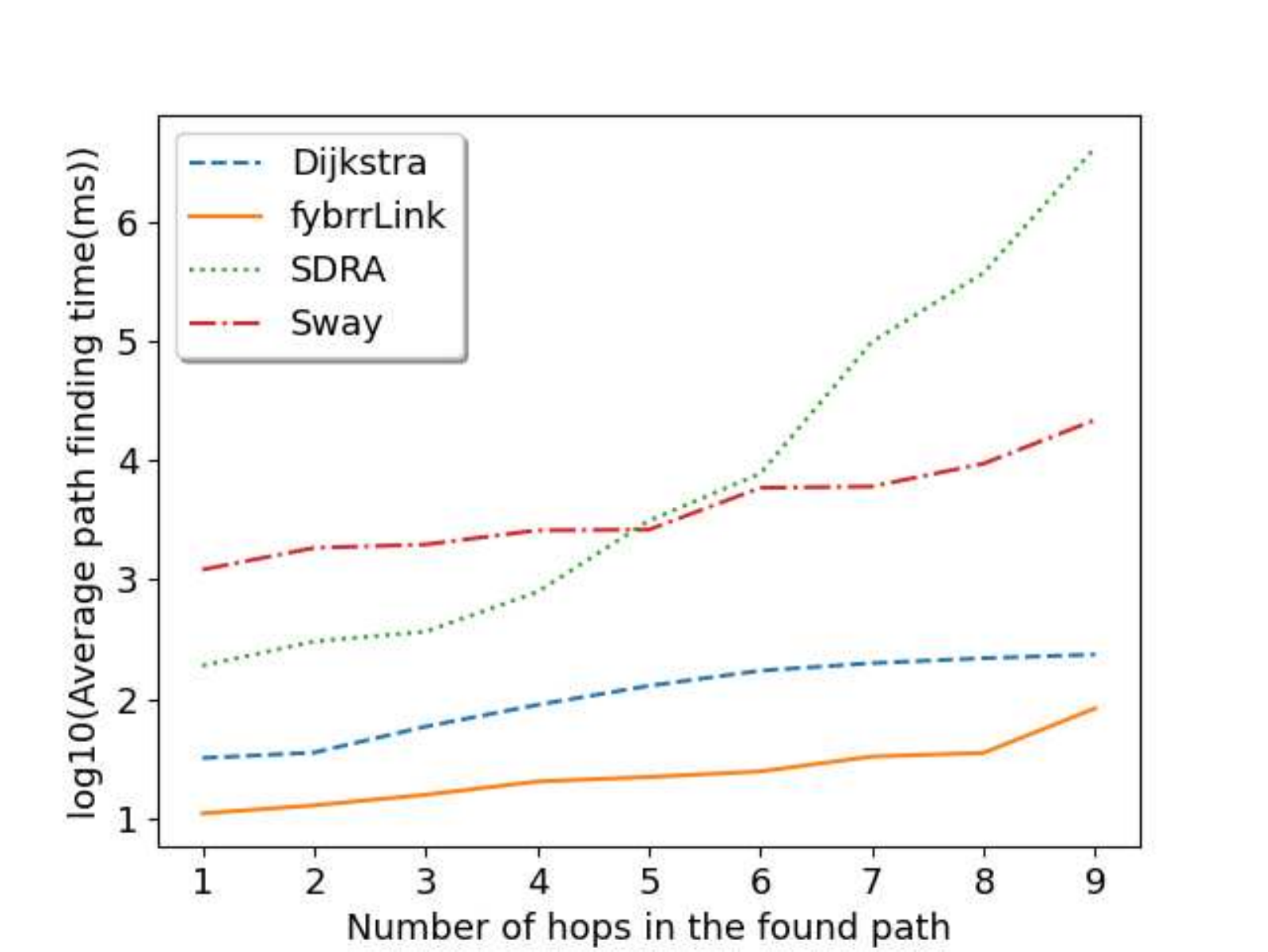}
    \caption{Average route finding time with increasing number of hops in routes.}
    \label{fig:avgexectimevshops}
\end{figure}

\begin{figure}
    \centering
    \includegraphics[width=8cm]{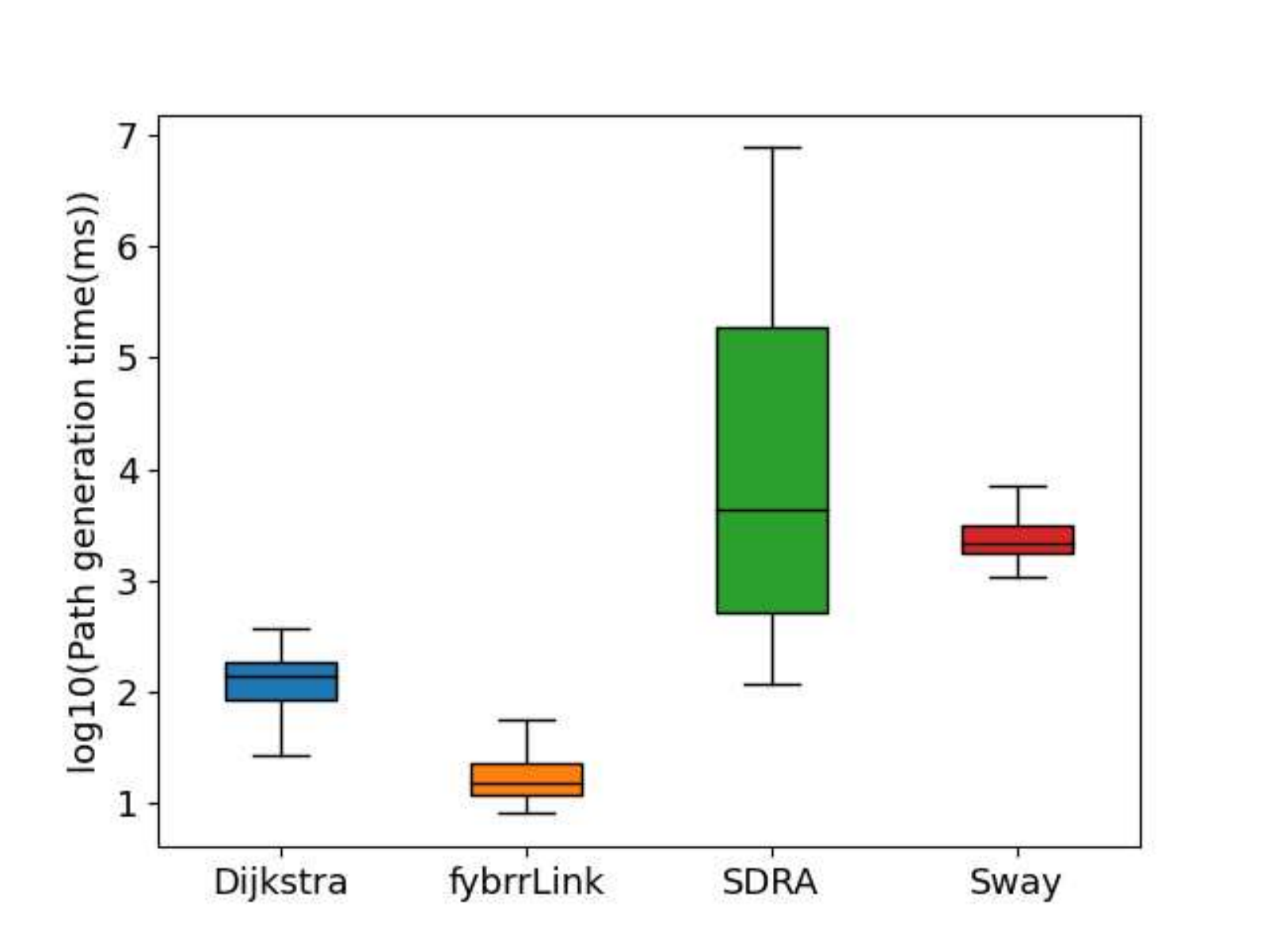}
    \caption{Route finding time of different schemes.}
    \label{fig:avgexectimebox}
\end{figure}
The variation of average flow route finding time with the number of hops in the generated path is illustrated in Fig. \ref{fig:avgexectimevshops}. Being a greedy algorithm, the route finding time of Dijkstra will increase with the number of hops in the resultant path. On the other hand, fybrrLink computes the ideal parallelogram to shorten the search space for our modified Dijkstra based on the coordinates of the source and destination node. As the number of hops increases in the resultant path, the size of the ideal parallelogram will also increase and that will lead to relatively bigger search space and route finding time will also increase with the increase in the number of hops. The average path finding time of Dijkstra is 3.4 times of fybrrLink. It is noteworthy that fybrrLink outperforms all other schemes in terms of route finding time with a very large difference. All the feasible paths are calculated in SDRA with the help of Depth First Search and this will take exponential time and thus very inefficient in terms of the number of computations required. Sway relies upon the Yen's k shortest path algorithm having a time complexity of $O(KV(E+V\log V))$ which is much more expensive than the complexity of fybrrLink. The average route finding time of Sway is 140 times of fybrrLink and SDRA is taking 110 times more time than of Sway. 

In Fig. \ref{fig:avgexectimebox}, minimization of route finding time in fybrrLink can be easily observed. The median value of route finding time for fybrrLink is 15 ms which is $89.20\%$, $99.65\%$ and $99.24\%$ lower than that of Dijkstra, SDRA and Sway having median values of 139 ms, 4341 ms, and 2127 ms respectively. Hence, fybrrLink is producing optimal paths with minimized route finding time.

\begin{figure}
    \centering
    \includegraphics[width=8cm]{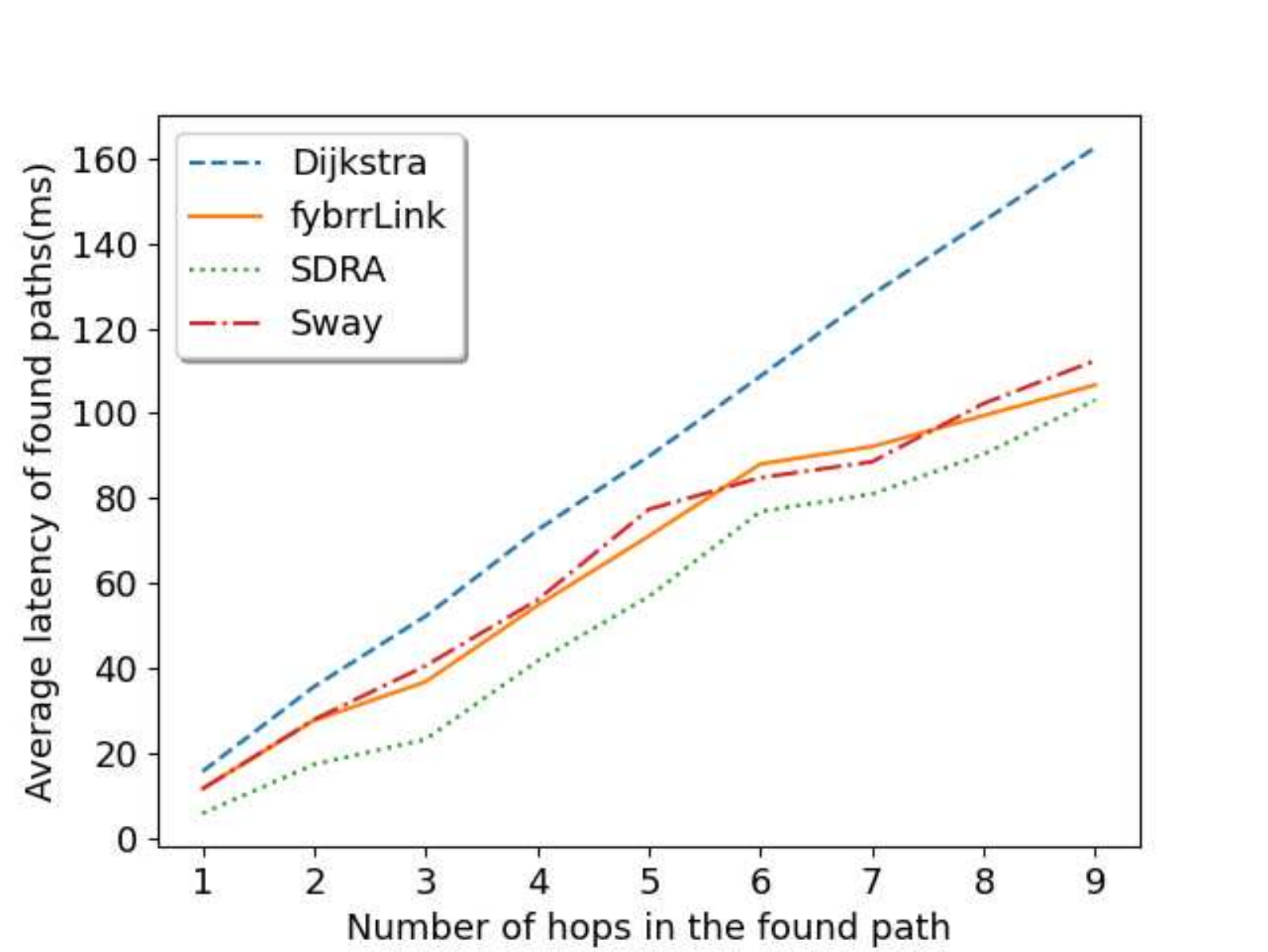}
    \caption{Average Latency of routes with increasing number of hops in route.}
    \label{fig:avglatvshops}
\end{figure}
Meanwhile, Fig. \ref{fig:avglatvshops} shows the variation of the average latency of generated paths with the number of hops attached with the path. Even when, we have included latency in the weights for Dijkstra, this graph shows fybrrLink generates the path with lesser average latency in comparison to the ones produced by Dijkstra. One reason for this is more queuing delay caused by the uneven distribution of flows over the various ISLs in Dijkstra. On the other hand, our algorithm ensures better distribution of load in the network. SDRA and Sway generate multiple paths (computationally expensive) and chooses the one with lesser latency. On average, paths generated by SDRA have $1.28\%$ lower latency than of fybrrLink but its route finding time is more and resource utilization is very poor. The average latency of routed flows using fybrrLink is $33.26\%$ and $0.34\%$ lower than that of Dijkstra and Sway, respectively.

\begin{figure}
    \centering
    \includegraphics[width=8cm]{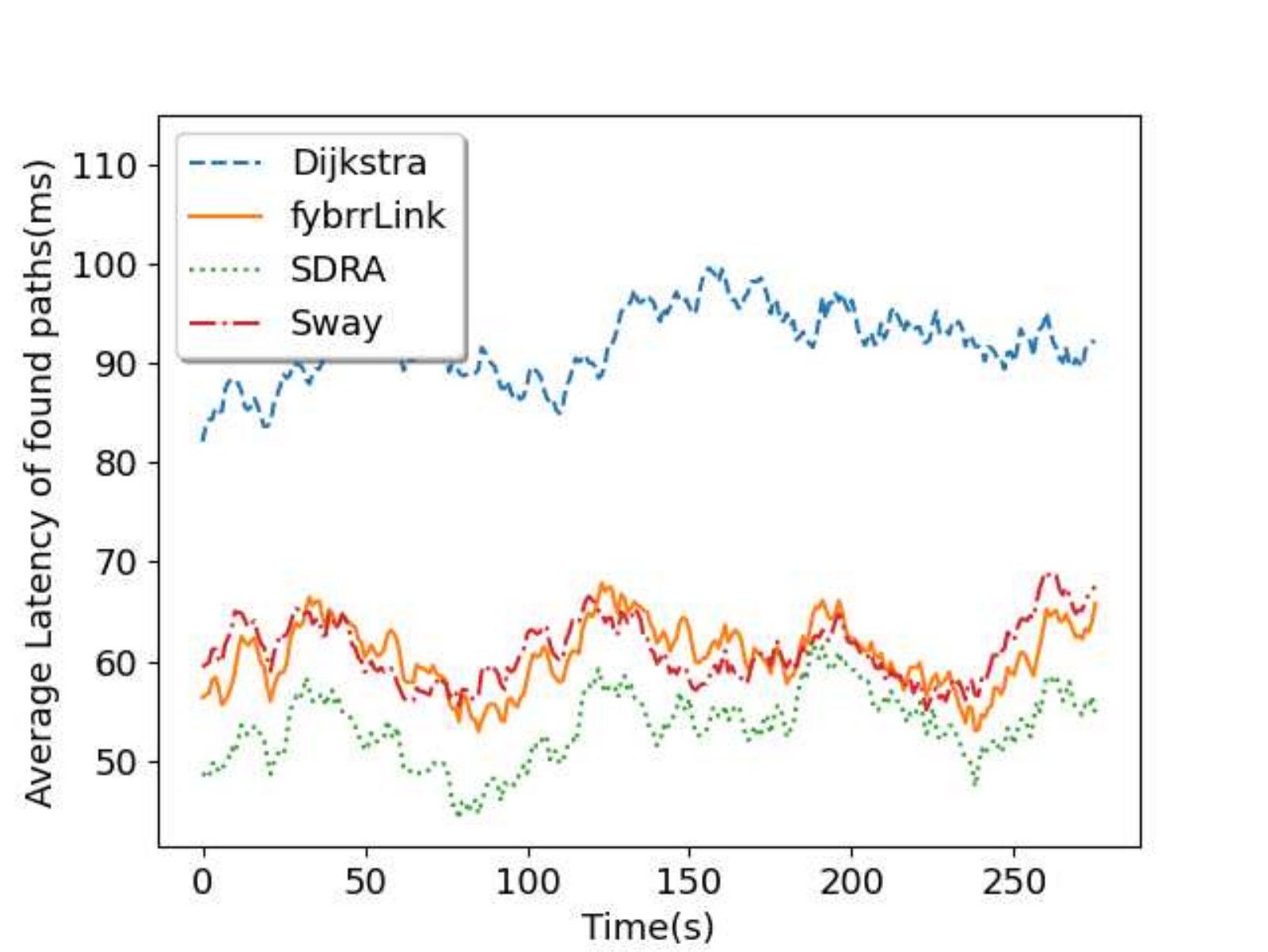}
    \caption{Variation of average latency of generated routes with time.}
    \label{fig:avglat}
\end{figure}

\begin{figure}
    \centering
    \includegraphics[width=8cm]{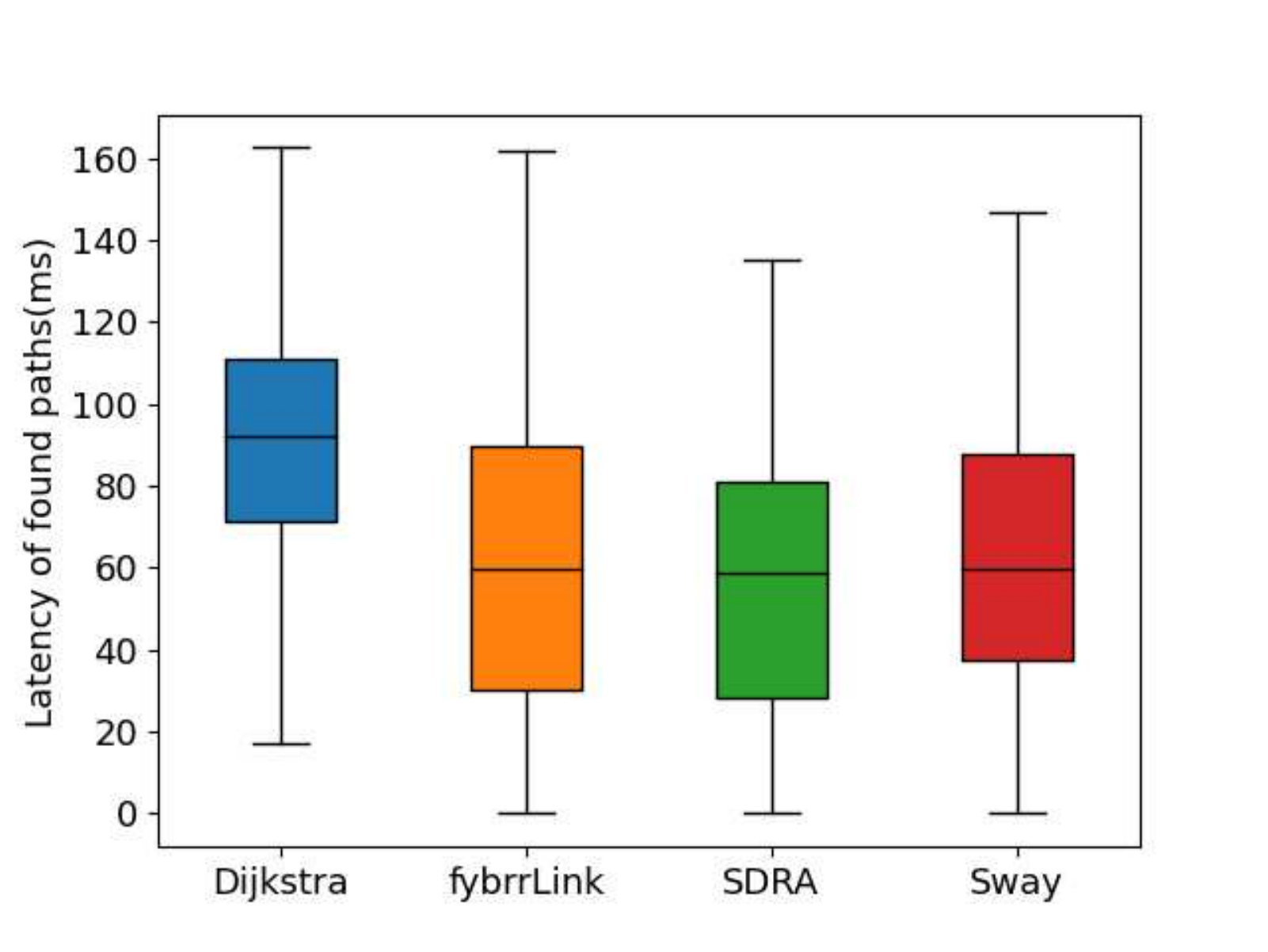}
    \caption{Latency of generated routes.}
    \label{fig:avglatbox}
\end{figure}

Fig. \ref{fig:avglat} and Fig. \ref{fig:avglatbox} compare the end-to-end latency of the generated paths of multiple schemes. From the beginning of the simulation to its end, the moving average (with window size 50 sec) value of latency is always higher in the case of Dijkstra because it is not choosing lower latency paths if the path is congested. The median value of path latency for fybrrLink is 59.77 ms which is 34.86$\%$ lower than that of Dijkstra having value of 91.77 ms. The median values of SDRA and Sway are 58.92 ms and 59.7703 ms  respectively. So, values of latency of generated paths are much comparable in the case of the other three algorithms but fybrrLink is getting the lower latency routes with a very low route finding time.

\begin{figure}
    \centering
    \includegraphics[width=8cm]{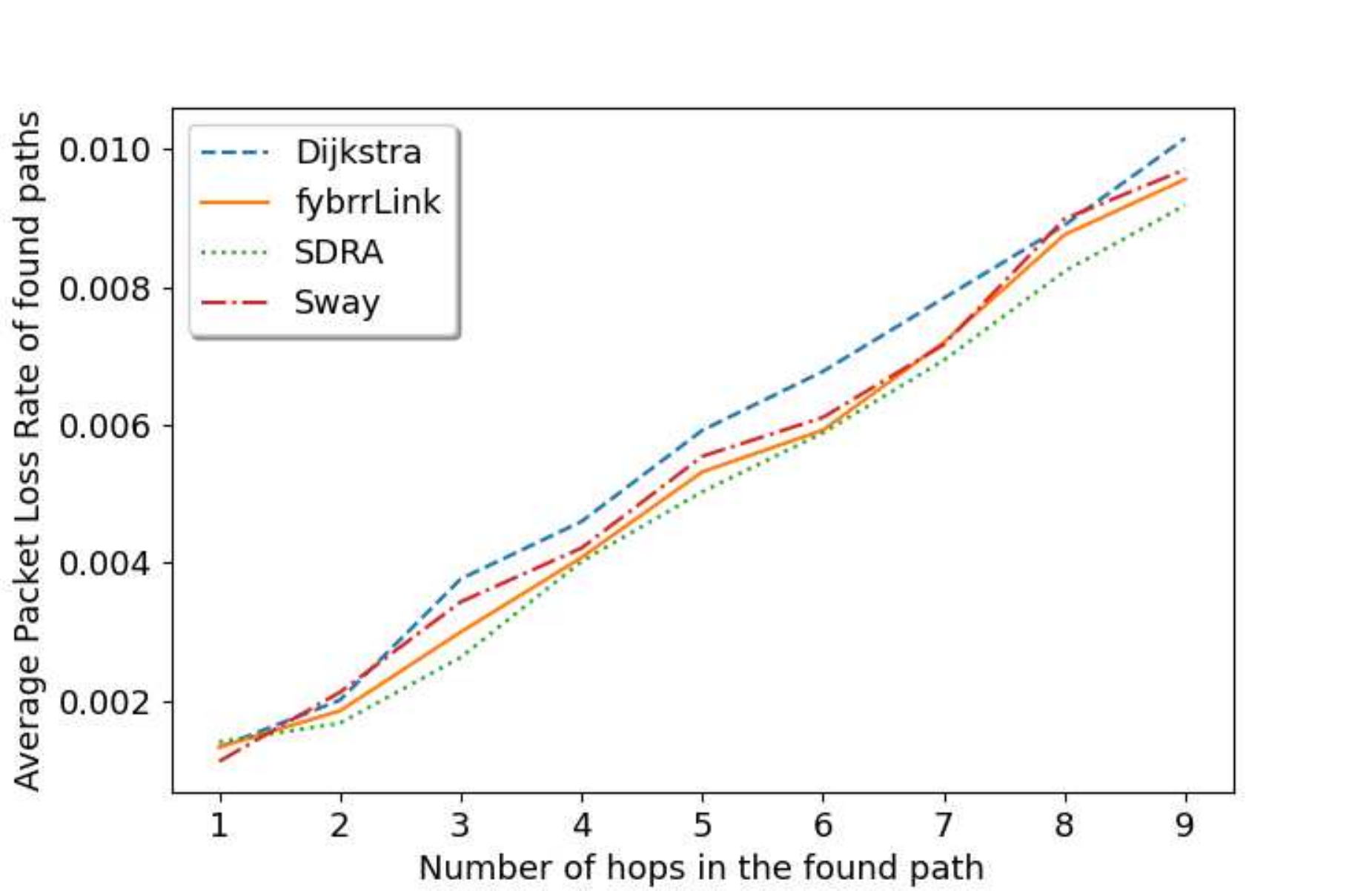}
    \caption{Average packet loss rate of routes with increasing number of hops in route.}
    \label{fig:avgPLRvshops}
\end{figure}

Fig. \ref{fig:avgPLRvshops} showcases the average packet loss rate values for different numbers of hops in the generated path. As we know, packet loss rate not only depends on the medium and number of hops but also on the congestion degree of the link. More congestion will lead to more packet loss and more queuing delays. fybrrLink includes the packet loss rate in its link scoring formula (See Eqn. \eqref{eqn:score}) and due to which congested ISLs are avoided in the final route by assigning them a larger weight, so resultant paths have much lower packet loss in comparison to other schemes. The average PLR of fybrrLink is 18.61$\%$ lower than that of Dijkstra. Sway does not takes PLR into the consideration so it performs a little poorly with this metric and its average PLR (over time) is 0.64$\%$ higher than that of our algorithm. SDRA choosing paths with the minimum number of hops will have lesser packet loss rates in generated paths but still, its average PLR is 14.31$\%$ higher than that of fybrrLink.

\begin{figure}[h]
    \centering
    \includegraphics[width=8cm]{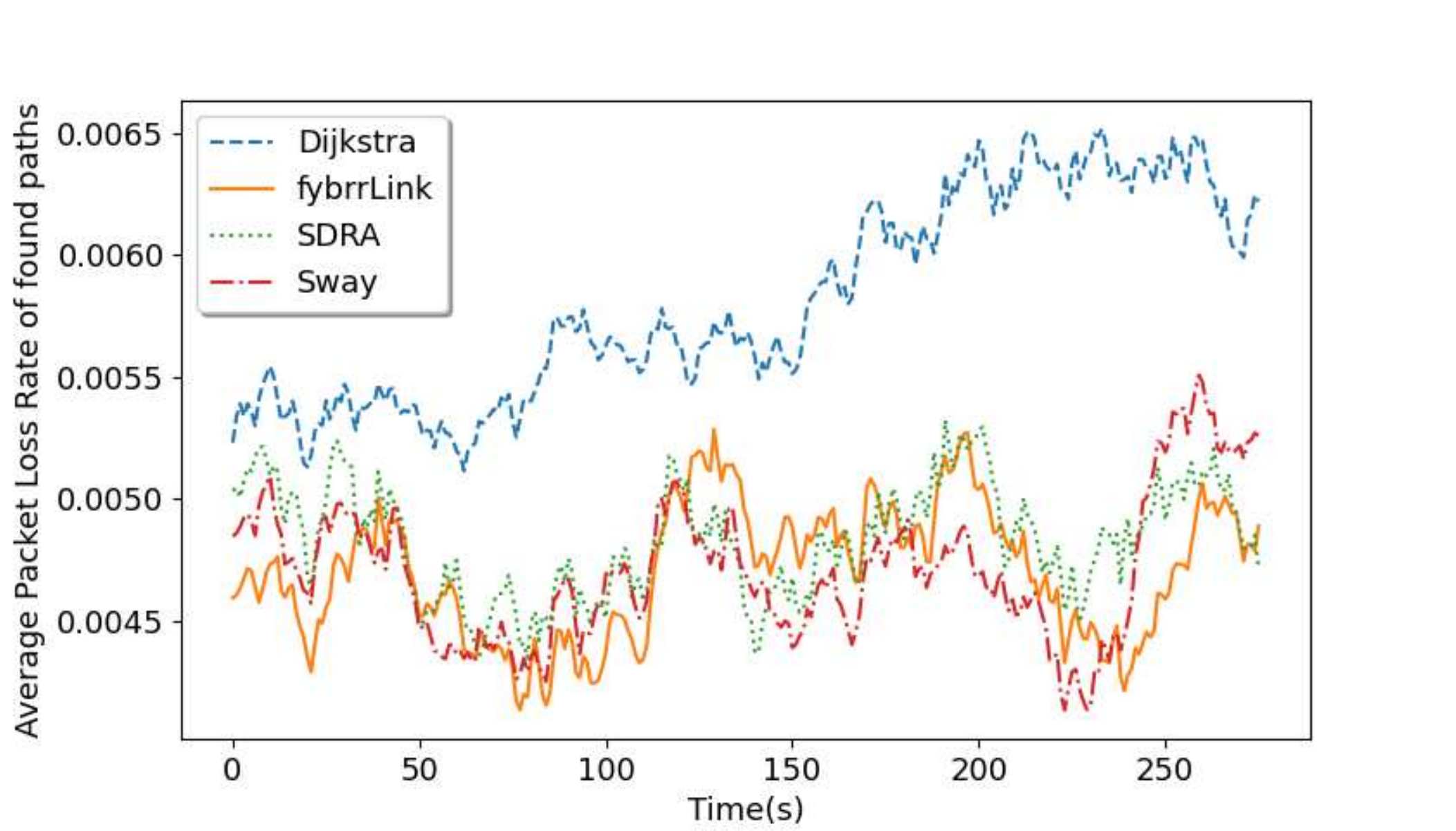}
    \caption{Variation of average packet loss rate of generated routes with time.}
    \label{fig:avgPLR}
\end{figure}

\begin{figure}[h]
    \centering
    \includegraphics[width=8cm]{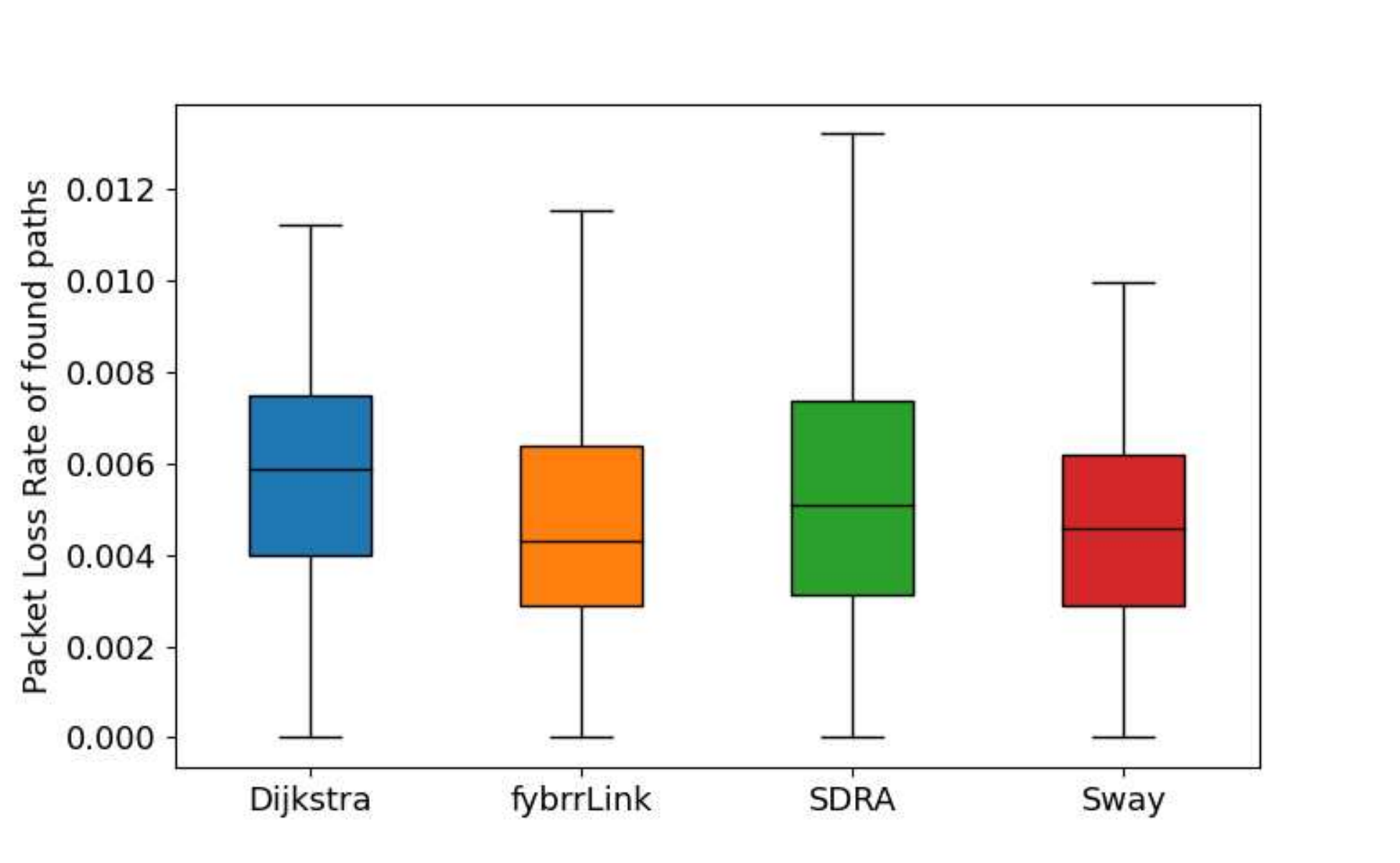}
    \caption{Packet loss rate of generated routes.}
    \label{fig:avglplrbox}
\end{figure}
Fig. \ref{fig:avgPLR} shows the variation of moving average value of packet loss rate with the simulation time and Fig. \ref{fig:avglplrbox} is to compare the same variation in packet loss rates with a box plot. Median value of packet loss rates for fybrrLink is $0.429\%$ which is $27.04\%$, $15.71\%$, and $6.53\%$ lower than that of Dijkstra, SDRA, and Sway having median values of $0.588\%$, $0.590\%$, and $0.459\%$, respectively. New flows were added every second, every new route is calculated based on the various factor of each link in the topology. The packet loss rate is one of those factors which our algorithm tries to minimize by assigning more weights to the links that are more prone to packet loss.

\setlength{\extrarowheight}{1pt}
\begin{table}[H]
 \centering
 \caption{Average values of various performance metrics for each scheme.}
 \label{table:results}
\begin{tabular}{|p{1.2cm}|p{2.6cm}|p{1.5cm}|p{1.2cm}|}
\hline
Scheme    & Routing time (ms)  & Latency (ms) & PLR     \\ \hline \hline
fybrrLink   & 37.76     & 60.74     &  0.468 $\%$                  \\ \hline
Dijkstra    & 135.93    & 91.01     &  0.575 $\%$                  \\ \hline
Sway        & 5321.08   & 60.96     &  0.471 $\%$                  \\ \hline
SDRA        &582124.18 & 59.46     &  0.535 $\%$                 \\ \hline
\end{tabular}
\end{table}
Table \ref{table:results}  summarises the average values of route finding time, latency, and packet loss rate for various schemes and clearly confirms the advantages gained by fybrrLink in terms of route finding time without any compromise with the quality of paths (See Fig. \ref{fig:dist_satisfaction} for details). Other schemes focus solely on reducing latency and congestion in the network. On the other hand, the inclusion of multiple other factors in the link scoring function allows fybrrLink to choose the most optimal paths with an efficient distribution of the network load over~ISLs.

The experimental results clearly show that fybrrLink is comparable if not better than the other compared schemes in various metrices. It is also able to satisfy QoS requirements of maximum flows with significantly lesser computation time and we achieved more available bandwidth, lesser latency, and a low packet loss rate in each of the routes calculated by our~algorithm.

\section{Conclusions}
\label{sec:conclusion}
In this paper, we have proposed and evaluated our QoS-aware routing algorithm for the network of LEO satellites. A maximum of 6 links is assigned to each satellite for routing and forwarding purposes. Global knowledge of SDN controllers about the network is employed to design \emph{fybrrLink}. Moreover, intuition and algorithms for flow transfer and topology monitoring are provided. Flow transfer is to minimize the impact of continuous topology alteration and topology monitoring is suggested to take advantage out of the predictable nature of satellite constellation. Further, NS3 simulation verifies the advantages of our proposed algorithms and approaches in terms of average route finding time, load distribution, average latency, and packet loss rate.

As future work, we would like to further explore the integration of terrestrial and non-terrestrial networks for efficient end-to-end data transmission. Controller placement problem can be studied in detail, the terrestrial controller can be compared with the controller at the non-terrestrial location, as this will also be a major factor for QoS-aware routing and optimized flow control. Security aspects of the NTN can also be studied and its effect on QoS-aware routing can also be considered.

\bibliographystyle{./bibliography/IEEEtran}
\bibliography{main}

\end{document}